\documentclass[
superscriptaddress,
amsmath,amssymb,
aps,
longbibliography,
]{revtex4-2}

\usepackage{graphicx}
\usepackage{dcolumn}
\usepackage{bm}
\usepackage{siunitx}
\usepackage{multirow}
\usepackage[version=4]{mhchem}
\usepackage{svg}
\usepackage{tikz}
\usetikzlibrary{calc}
\usepackage{tikz}
\usepackage{tikz-3dplot}
\usepackage{color, colortbl}
\usetikzlibrary{angles, quotes}

\usepackage[utf8]{inputenc}
\usepackage{xr}

\begin{document}


\title{Geometry-enhanced graph neural network for learning the smoothness of glassy dynamics from static structure}
\author{Xiao Jiang}
\affiliation{College of Computer Science and Electronic Engineering, Hunan University, Changsha, China}

\author{Zean Tian}
\affiliation{College of Computer Science and Electronic Engineering, Hunan University, Changsha, China}

\author{Kenli Li}
\affiliation{College of Computer Science and Electronic Engineering, Hunan University, Changsha, China}

\date{\today}

\begin{abstract}

Modeling the dynamic processes of glassy systems has been challenging in physics science for several decades. Recent studies have shown that graph neural networks (GNNs), which learn particle dynamics from the graph of glassy systems, can effectively capture correlations between the structure and dynamics. In contrast to these unprincipled approaches, this paper introduces a novel dynamical parameter, smoothness, to establish the relationship between structure and dynamics from the theory of graph signal processing. The observation of smoothness motivates us to encode the structural features under dynamic constraints. However, current graph-based approaches ignore these constraints, leading to a decline in the correlation between structure and dynamics, especially in the case of a short timescale. Therefore, we propose a geometry-enhanced graph neural network (Geo-GNN) to learn the smoothness of dynamics. Experimental results demonstrate that our method outperforms state-of-the-art baselines in predicting the dynamics over all time scales examined. Ablation experiments reveal the effectiveness of each proposed component in capturing smoothness in dynamics. This finding not only refines the method for predicting glassy dynamics but also provides the lens through which to investigate the issue of causality regarding dynamical heterogeneity in the glassy system.

\end{abstract}

\maketitle


\section*{Introduction}

Glass is a prototypical example of non-equilibrium matter, formed when a fluid is quenched or densified until it ceases to flow \cite{angell1995formation,ediger1996supercooled, debenedetti2001supercooled, randomfirstorder,berthier2011theoretical, keys2013calorimetric}. In contrast to more ordinary condensed matter, this solidification shown in the glass transition occurs without any discernible structural order seen in conventional correlation functions \cite{berthier2011theoretical}. However, minor temperature variations can significantly vary viscosity and structural relaxation time by many orders of magnitude. Although such glass translation phenomena can be observed in various systems with dynamic correlation sizes ranging from microscopic to macroscopic \cite{angelini2011glass}, its mechanism remains a mystery. 

One of the hallmarks of the glass transition is dynamic heterogeneity, where particles in glassy supercooled liquids are clustered together in the form of cooperatively rearranging regions that move faster or slower than the average \cite{ediger2000spatially}. The concept of dynamic heterogeneity has been at the forefront of research on glass and non-equilibrium systems for the past two decades. Numerous attempts have been made to establish a structural signature of dynamics in supercooled liquids. Due to the lack of universal structural order, traditional approaches \cite{leocmach2012roles, malins2013identification, tong2018revealing, tong2019structural, marin2020tetrahedrality, widmer2006predicting, widmer2008irreversible} provide weak structure-dynamics correlations that are however insufficient for constructing a theory. 

Recent advances in high-performance technology and artificial intelligence (AI) \cite{li2014performance, zeng2021efficient,9947316} have facilitated the application of machine learning (ML) algorithms in various fields \cite{duan2020novel,clegg2021characterising,liu2019machine}. In particular, machine learning offers a new avenue for identifying the structural predictors associated with dynamic heterogeneity. The first application of support vector machines (SVMs) to the dynamics prediction of glassy systems discovered that machine-learned softness could identify slow and fast regions in liquid glass \cite{cubuk2015identifying,ma2019heterogeneous,schoenholz2016structural}. After that, various techniques have been proposed for multiple glassy systems, from supervised approaches like linear regression and graph neural network (GNN) \cite{BAPST, PaperEmanuele, wang2021inverse, shiba2023botan, jung2022predicting, pezzicoli2022se3equivariant} to unsupervised methods like community detection and auto-encoder \cite{structurevsdynamics5, boattini2020autonomously, bihani2022unsupervised, coslovich2022dimensionality, oyama2022deep}. In particular, GNN has achieved unprecedented accuracy in predicting the dynamic propensity by only using the static graph structure of a glassy system as input \cite{BAPST,shiba2023botan,pezzicoli2022se3equivariant}. For example, Ref. \cite{pezzicoli2022se3equivariant} proposes a rotation equivariant GNN (SE(3)-GNN) model from the inductive bias of symmetry for learning the structure representation of the glassy systems. More recently, several works attempt to improve the prediction of glassy dynamics with physical intuition, for example, by informing the cage state of the initial configuration \cite{alkemade2023improving}, or using the physics-informed input extract from the inherent structure \cite{jung2022predicting}, or by adding the relative motion between particle pairs \cite{shiba2023botan}. Although absorbing additional dynamic information in these methods brings comparable performance when predicting dynamic propensity, a powerful prediction model solely from the initial configuration is still desired. Additionally, even though these methods answer from different perspectives what was left out when trying to predict the dynamics, the question of what kind of structure affects the dynamics remains explicitly unresolved.

{Here, we find that the current graph-based approaches ignore the smoothness constraints between node targets. The learned structure will make the predicted dynamical field a smooth system in which spatially neighboring particles tend to have similar dynamics.} For example, ref. \cite{PaperEmanuele} simplified the feature encoder of the GNN \cite{BAPST} with intelligent structure descriptors based on message-passing rules. This strategy involves two steps: feature descriptor construction and feature aggregation. In the descriptor construction step, the structure around the particles is encoded by hand-crafted features. The second step can be considered as the feature aggregation on the graph, which is achieved by shell averaging. Then, the aggregated features can be fitted to the dynamics of multiple time scales. It was shown that linear regression can predict as well as the GNN when given such smart information \cite{alkemade2022comparing}. {The shell averaging is of great importance for success in predicting dynamics because averaging the local structure can optimize the feature to match the underlying smooth constraint, i.e., particles with similar propensities tend to be spatially adjacent.} However, the dynamics between neighbors are sometimes non-smooth, i.e., particles with different propensities tend to be adjacent. The caging dynamics is a typical non-smooth system in which the fast-moving particles are always connected to the slow particles due to the cage effect. {In this case, shell averaging causes the nearby particles to have similar structural features, which makes it impossible to distinguish the true dynamics of the fast atoms from the slow ones of their nearby neighbors. This suggests that the shell averaging used by current methods for predicting glassy dynamics is far from optimal. That is, not all the structure-dynamics correlations benefit from the akin embeddings between connected particles. So, we need to find correlations between structure and dynamics by considering the underlying dynamic pattern triggered by the near-neighbor particles.}

This paper introduces the dynamic smoothness of the glassy system, a novel concept that can depict the spatial relationship of dynamics from a graph perspective. We start by leveraging the graph signal processing tools to review the glassy dynamics on the graph, both globally and locally. The graph analysis results reveal that the dynamics at different time scales exhibit varying degrees of smoothness. Keeping this concept in mind, we propose two types of frameworks to enhance the capability of learning the smoothness of dynamics. Based on these frameworks, we propose a novel geometry-enhanced Graph Neural Network (Geo-GNN) that can efficiently predict the dynamics on different time scales from a single static structure. Specifically, we meticulously developed the Geo-GNN encoding and aggregation processes to provide a more accurate depiction of underlying particle smoothness. Experimental results on the Kob-Andersen system show that our model can autonomously understand the underlying dynamic patterns and surpass the existing approaches in predicting short- and long-time dynamics. This is the first report on graph theory to study structure-dynamic relationships in glass systems. 

\section*{Results}
\textbf{The molecular dynamic simulation data set} 

In this research, the molecular dynamics simulation data for an 80:20 Kob-Andersen (KA) system is adopted to evaluate the predictive power of our model. The KA system is a mixture of $N = 4,096$ particles: $N_A = 3277$ (large) Lennard-Jones (LJ) particles of type $A$  and $N_B = 819$ (small) LJ particles of type $B$. The dimensionless units based on the LJ potential are used here. To give a fair comparison, we used the data released in Ref.~\cite{BAPST}. More detailed information on simulations can be found in their original paper. The focus of the state point here is on the temperature, $T = 0.44$, which is close to the transition temperature predicted by Mode Coupling Theory\cite{donati1999spatial}, $T_{MCT} = 0.43$. The built data set comprises 800 independent configurations, of which half are used for training and the rest for testing. For each equilibrated configuration, the dynamic propensities of large particles are computed from 30 trajectories with identical beginning coordinates but random velocities selected from a Maxwell-Boltzmann distribution \cite{widmer2004reproducible, propensity1}. Formally, the dynamic propensity of particle $i$ at time $t$ is defined as $\Delta_{i}(t) = \langle|\bm{r}_{i}(t) - \bm{r}_{i}(0)|\rangle$, where $\langle..\rangle$ represents the ensemble average and $\bm{r}_{i}(t)$ represents the position vector of particle $i$ after a time interval ${t}$. Particles with large propensity move faster.

Dynamic propensity serves as the target for machine learning to identify relationships between local structure and dynamics. With the static structure of the equilibrated configurations, our objective is to predict the propensities of type $A$ particles at various time scales from the static structure. Throughout the text, the real part of the self-intermediate scattering function $S(\bm{q},t) = 1/N_A\langle\sum_{j=1}^{N_A}cos(\bm{q}(\bm{r}_{j}(t) - \bm{r}_{j}(0)))\rangle$, where $\bm{q}$ is the wave vector corresponding to the first peak of the static structure factor, is used to represent the time $t$ for simplification. The structural relaxation time $\tau_g$ corresponds to $S(\bm{q},t)=1/e$. 

\begin{figure}[htbp]
  \centering
  \includegraphics{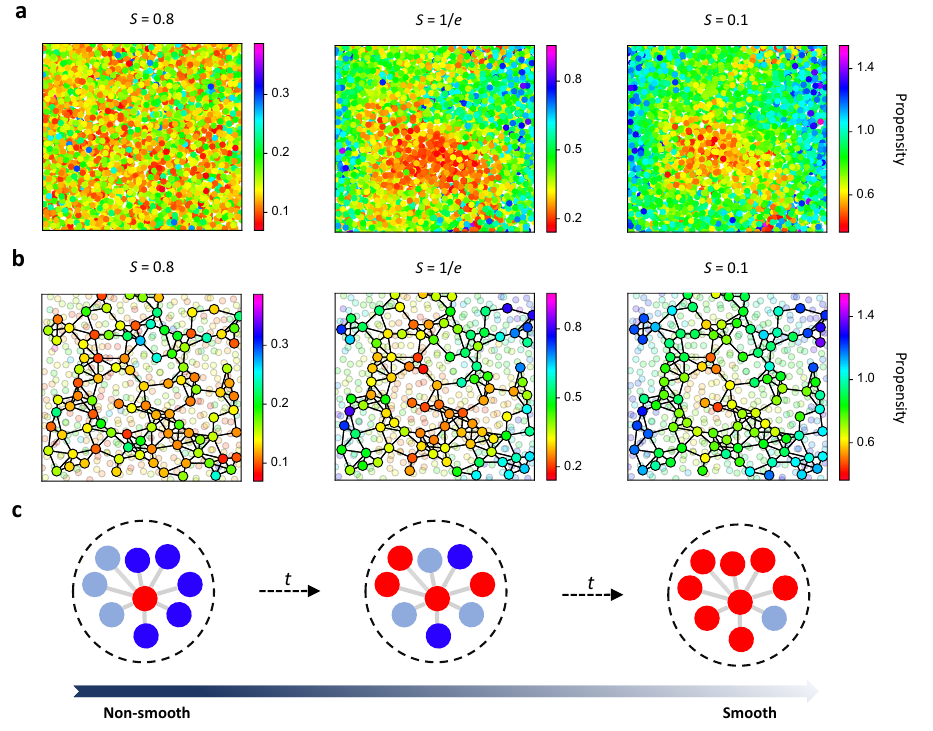}
  \caption{\textbf{Particle dynamics at different timescales} \textbf{(a)} Top view of a configuration with each particle colored with its dynamic propensities at different time scales. \textbf{(b)} The dynamic propensities at different time scales are assigned as the graph signals in the graph representation of the configuration. \textbf{(c)} An example to illustrate the notion of smoothness is from the view of the local graph. When most of the graph signals of the neighboring nodes are the same color as the graph signals of the central nodes (red), the graph gradually becomes smooth.}
  \label{fig:example}
\end{figure}

\textbf{The smoothness of the dynamics}

As is well-known, the behavior of the dynamics in the caging regime differs significantly from that in the diffusive regime. In the short-time scale, where $S(\bm{q}, t) =0.8$, particles perform vibratory movements in a temporary cage produced by their neighbors, exhibiting a low spatial correlation of the motion. In contrast, in the long-time scales where $S(\bm{q}, t) = 1/e$ and $0.1$, the spatial correlation of their motion increases as the cage effect fails, and particles move cooperatively. Fig.~\ref{fig:example}(a) illustrates a top view of one configuration in which the particles are colored according to their propensity for three distinct time scales. The spatial correlations of the mobility field differ markedly between short- and long-time dynamics. At longer timescale, particles with similar dynamic propensities tend to cluster together \cite{PhysRevLett.116.085701,tah2020signature}, resulting in a smoother system appearance. However, at shorter timescales, the system appears less smooth, with the dynamics of spatially adjacent particles behaving differently. Instead of quantifying the dynamic behavior only from the dynamics themselves, we examine this apparent dynamic behavior from a graph perspective. Inspired by the emerging field of graph signal processing \cite{shuman2013emerging}, we consider the dynamic propensities of particles as signals on the graph. The spatial correlation of particle dynamics can then be explored through the smoothness of graph signals. Here, we understand the dynamics of the graph by quantifying both the global smoothness from the global graph perspective and the local smoothness from the local perspective (see Methods).

Given a three-dimensional input for a configuration,  we first construct an undirected graph for larger particles based on the cutoff distance $R_c = 2$ (as shown in Fig.~\ref{fig:example}(b)). We can see that the dynamics in the short-time scale, where $S(\bm{q}, t) =0.8$, fluctuate more strongly among connected nodes than those in the long-timescale, where $S(\bm{q}, t) =1/e, 0.1 $. To better illustrate this, we use the Rayleigh quotient to explore the global smoothness of a configuration at different timescales. Specifically, with the Laplacian graph matrix $\bm{L}$ and signal $\bm{x}$ supported on the graph, the Rayleigh quotient $\frac{\bm{x}^T\bm{L}\bm{x}}{\bm{x}^T\bm{x}}$ quantifies the smoothness of $\bm{x}$ by measuring the changes of the signal along the nodes. The lower the Rayleigh quotient, the slower the graph signals vary on the graph. For a graph built from a configuration in which the dynamics propensities at different timescales act as a graph signal $\bm{x}$, we can obtain the Rayleigh quotient values as a function of time.

\begin{figure}[htbp]
  \centering
  \includegraphics{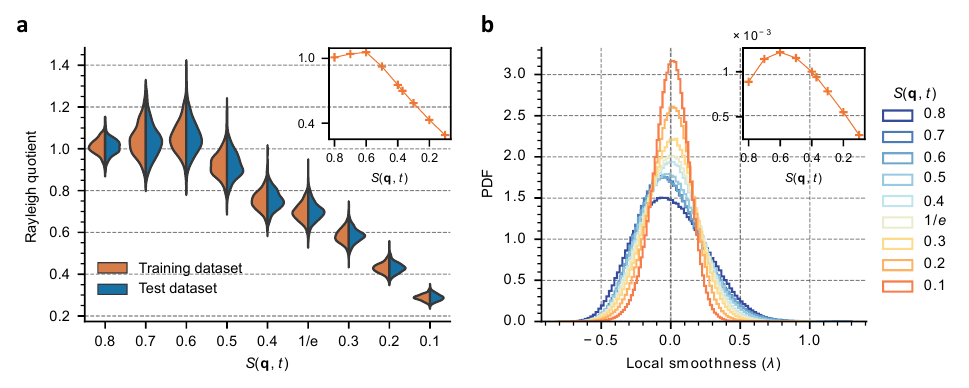}
    \caption{\textbf{Global and local smoothness of dynamics} \textbf{(a)} Violin plot of Rayleigh quotient as a function of time at temperature $T = 0.44$. The insert shows the mean values of the Rayleigh quotient. \textbf{(b)} The probability distribution function (PDF) of the local smoothness $\lambda$. The insert shows the mean values of $\lambda$. }
  \label{fig:g-l}
\end{figure}

Fig.~\ref{fig:g-l}(a) shows the global smoothness of particle dynamics, as expressed by the Rayleigh quotient of all training and test graphs. The insert displays the mean value of the Rayleigh quotient obtained by averaging over all graphs. The distributions of the Rayleigh quotient in the training and test graphs are nearly identical. The Rayleigh quotient is highest at short timescales where $S(\bm{q},t)$ is approximately 0.8, 0.7, and 0.6, indicating that dynamic propensities fluctuate dramatically between graph nodes and show a high-frequency dynamic pattern on the graph. On a longer time scale, the Rayleigh quotient decreases over time, indicating a low-frequency pattern of dynamics. This is consistent with the spatial correlations of particle dynamics shown in Fig.~\ref{fig:example}(a). For a short period, particles perform vibratory movements in individual cages produced by their neighbors, resulting in a low spatial correlation of the motion. Therefore, the propensities of neighboring nodes on the graph tend to differ on shorter timescales, represented by a high Rayleigh quotient. When the cage effect fails, particles move coordinately, leading to the formation of clusters. The spatial correlation of their motion grows, and the dynamics at the neighboring nodes tend to be similar on longer timescales, as shown by a low Rayleigh quotient.

To further investigate the local smoothness of dynamics at the particle level, we computed the local smoothness $\lambda_{i}$ between each central particle and its first nearest neighbors (see Methods). A larger absolute value of $\lambda_{i}$ represents the dynamic of the particle $i$ is non-smooth, whose dynamic propensity differs significantly from its first nearest neighbors on the graph (as shown in Fig.~\ref{fig:example}(c)). Fig.~\ref{fig:g-l}(b) describes the distribution of the local smoothness for particles in all training and test graphs. As time progresses, the distribution becomes centered around zero. We observe that non-smooth particles are prevalent in shorter-time dynamics, whereas the number of smooth particles increases in longer ones. However, non-smooth particles still exist even in longer-time dynamics. The insert of Fig.~\ref{fig:g-l}(b) displays the mean value of local smoothness obtained by averaging over all particles. The trend of local smoothness over time is consistent with global smoothness, indicating that both global and local smoothness can demonstrate the inherent smoothness within particle dynamics. 

\textbf{Smoothness-constrained structure representation}

Now the question is, what kind of structural features do we need under the smoothness constraint of dynamics? The answer to the question is also straightforward: the structural features should correspond to the smoothness pattern of the dynamics. Unfortunately, current models struggle to handle smoothness attribution. For example, neighboring particles can exhibit distinct dynamic propensities on short timescales and form a non-smooth pattern. If we attempt to average the local structure \cite{PaperEmanuele}, which acts as a low-pass filter, the differences between structural features (high-frequency information) of linked particles will be masked. In contrast, the commonalities (high-frequency information) will be preserved. Finally, the obtained structural representation will force to be identical. Eventually, the classifier will erroneously group particles with distant dynamics into one category. We can find that averaging local structure often leads to poor predictions for non-smooth dynamics \cite{PaperEmanuele}, because the low-frequency information is no longer sufficient to infer non-smooth patterns of dynamics. Therefore, we also need to look for high-frequency structure representation. 
\begin{figure}[htb]
  \centering
  \includegraphics{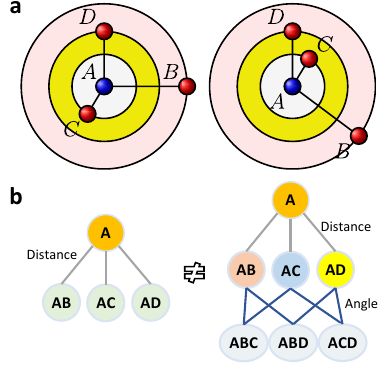}
  \caption{ \textbf{(a)} A pair of local structures of center particle $A$ has different shapes but with the same list of distances. \textbf{(b)} The structures cannot be discriminated by only using the unordered list of distances but can be classified with additional angle information. }
  \label{fig:angle}
\end{figure}

As illustrated above, one can realize that the inherent smoothness of glassy systems is crucial to guide our design of structural features for predicting particle dynamics. Then why can not the current models learn high-frequency features for non-smooth dynamics?

One of the reasons is due to the limited geometric expressive power. Specifically, in graph deep learning, researchers have demonstrated that the expressive potential of GNN with message-passing architectures is upper bounded by the 1-WL (first-order Weisfeiler-Lehman) graph isomorphism test\cite{xupowerful}. As for particles in the three dimensions, the 1-WL test on the distances graph with finite cutoff has also demonstrated the limited expressive power of GNNs using only distances \cite{pozdnyakov2022incompleteness}. For example, Fig.~\ref{fig:angle} shows two distant local environments of central particle $A$. Using only the unordered list of distance, it is not possible to distinguish between two distantly positioned particles. Consequently, GNNs will map them to the same points in the embedding space. The structural differences between neighboring particles and even all particles in the system will be restricted, rendering them indistinguishable and affecting the prediction of the dynamics. However, we can re-separate them in the feature space by adding additional structural information, such as angles, as shown in Fig.~\ref{fig:angle}(b). Thus, we can improve the expressive power of the geometric structure to learn the high-frequency pattern, creating high-dimensional features that reflect differences between neighboring particles. In combination with these high-frequency features, it will be easier to distinguish between different glass dynamics. The recently proposed SE(3)-GNN is representative of this class of methods. The equivariant representations of glassy in SE(3)-GNN can characterize the local structure more abundantly than GNN. Thus, effective high-frequency features can be learned and improve predictive power.

Another reason is the shortcomings of the GNN aggregation strategy itself. Researchers have pointed out that GNNs work like low-pass filters, smoothing differences between neighboring nodes by removing high-frequency signals \cite{nt2019revisiting, zhou2020graph}. As a result, the learned representation of central particles will converge to the averages of their local neighbors. Shell averaging proposed in \cite{PaperEmanuele} is a typical example. This may be sufficient for predicting glassy dynamics when the spatial correlation scale of dynamic propensities is large since smoothness dominates and fast or slow particles tend to form patches. However, when the spatial correlation length scale decreases, predicting non-smooth patterns of dynamics becomes more challenging. Thus, the final features used for predicting dynamics should be aggregated based on the actual smoothness of dynamics. Simple shell averaging is only effective for smooth particles, so we need an adaptive aggregation approach to handle varying degrees of smoothness. Learning the relative motion between particles is representative of the improved aggregation approach \cite{shiba2023botan}. The authors have added the task of learning relative motion on the last layer of aggregation of GNN as an explicit constraint on the local dynamics. The structural features between neighbor particles can be prevented from over-smoothing by learning this additional dynamic information. Thus, It can also learn a high-frequency representation and effectively predict short-time dynamics. In contrast to the approach taken in this paper, we aim to avoid explicitly using additional dynamics and instead learn the smoothness of the dynamics by adding implicit constraints.

In conclusion, we point out two ways to improve the ability to learn the underlying smoothness of dynamics. One way is to enhance the geometric expressive power of the model. It can prevent the model from learning only low-frequency structural features and thus provide reasonable embedding space to further exploit the structure-dynamic relationship. Another one is to improve the aggregation strategy of GNNs so that the aggregation can be conducted under the smoothness constraints of the dynamics. By considering these factors, we can enhance the ability to handle smoothness patterns of dynamics.

\begin{figure}[htb]
  \centering
  \includegraphics{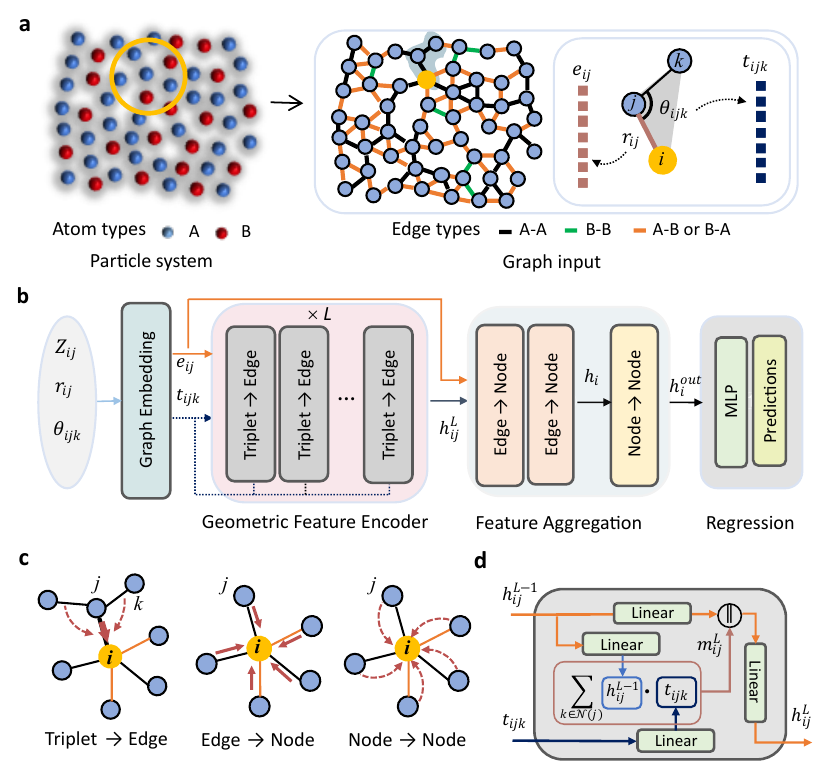}
  \caption{\textbf{Illustration of the geometry-enhanced graph neural network (Geo-GNN)}. \textbf{(a)} A three-dimensional glass system is represented by a graph, where particles are taken as nodes and two nodes at a distance that is less than 2 LJ units will be connected by an edge. The types and distances are assigned as the edge feature. Angles corresponding to triplet particles are assigned to two connected edges. \textbf{(b)} Workflow of Geo-GNN: graph embedding, geometric feature encoder, feature aggregation, and regression. \textbf{(c)} Message passing scheme for triplet-to-edge layer, edge-to-node layer, and node-to-node layer. \textbf{(d)} Construction of the triplet-to-edge layer.}
  \label{fig:model}
\end{figure}

\textbf{The Geo-GNN model} 

The proposed architecture is detailed in Fig.~\ref{fig:model}. The initial features of type, distance, and angle are added to each constructed graph for the Geo-GNN. An embedding layer processes these details to yield representations for edges and triplets. The stack of triplet-to-edge interaction layers propagates edge and triplet information, giving the learned local structural features at each edge. Edge-to-node and node-to-node interaction operations are performed to aggregate the features of each particle in the local sale, followed by multi-layer perception (MLP) for the final predictions of particle propensity. Further details on the model and training process can be found in Methods.

To handle the smoothness of dynamics, we follow the above instructions to develop the model architecture to learn high-frequency representations. Precisely, the model consists of the following two main parts. 

(1) Firstly, the Geo-GNN leverages a geometric-informed high-order message passing for describing local environments. To boost the geometric expressive power,  the three-body connections and the corresponding angle features are used in the triplet-to-edge layers of Geo-GNN. The improvement of structural expressiveness allows the model to differentiate more local environments than 1-WL message passing and thus bring more high-frequency patterns of structural features between neighboring particles. 

(2) Next, within a more recognizable structure space, the final features used to predict the dynamics are further aggregated according to the dynamical smoothness. Instead of averaging, we develop an adaptive aggregation process with a self-attention mechanism to handle varying degrees of smoothness of particle dynamics. With attention on the edge, the features of neighboring particles with identical targets will combine to produce low-frequency dynamics patterns. On the other hand, neighboring particles with distant dynamics will drive away in the feature space to learn non-smooth dynamics. Thus, we can learn the structural features of the non-smooth or smooth targets from neighbor particles in an implicit way. By incorporating these techniques, the Geo-GNN model shows a high prediction accuracy of particle dynamics against the current model.
\begin{figure}[htb]
  \centering
  \includegraphics{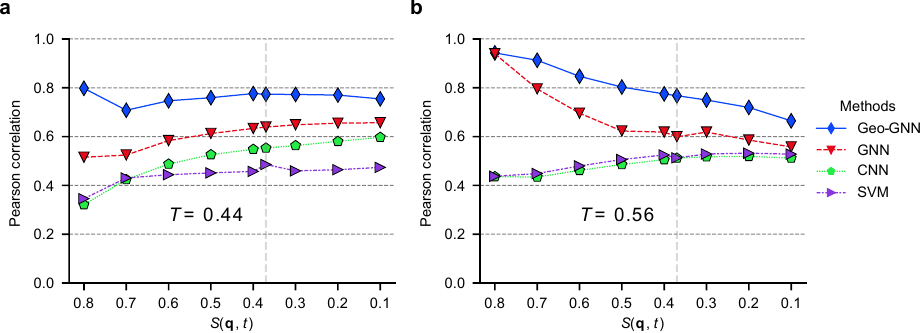}
  \caption{ \textbf{Comparison with GNN of predicting particle propensity in the KA system}. Pearson correlation coefficient between predicted and actual propensities for $A$ particles in the KA system at temperature \textbf{(a)} $T = 0.44$ and \textbf{(b)} $T = 0.56$, respectively. The vertical gray dashed lines indicate the timescale $t = \tau_g$, corresponding to $S(\bm{q}, t) = 1/e$. The results of support vector machine (SVM), convolution neural network (CNN), and graph neural network (GNN) are taken from Ref. \cite{BAPST}. }
  \label{fig:re1}
\end{figure}

\textbf{Comparison to GNN}

We first evaluate the performance of the proposed Geo-GNN in predicting dynamic propensities using the Pearson correlation value between the predicted and actual propensities. A higher correlation coefficient indicates better performance. Fig.~\ref{fig:re1} shows the comparison results at the temperatures $T = 0.44$ and $T = 0.56$.

When compared to the state-of-the-art GNN, it is noticeable that our approach shows significantly better prediction performance. At the lowest temperature $T = 0.44$, our model shows overwhelming superiority over GNN in predicting the short-time dynamic propensities, specifically from the caging regime where $S(\bm{q}, t) = 0.8$ to the relaxation time where $S(\bm{q}, t) = 1/e$. Especially in the caging regime, Geo-CNN achieves the best performance, while the prediction at these times seems to be the most challenging for GNN. This suggests that Geo-GNN can capture more structural factors for dynamic predictions than GNN. Although the performance of GNN improves after the relaxation time, our model remains much superior.

\textbf{Comparison to the models of improved geometric expressive power}

Next, we compare with other recently proposed models in \cite{PaperEmanuele}, \cite{pezzicoli2022se3equivariant}, and \cite{jung2022predicting}. These works are also designed to enhance the geometric expressiveness power. The linear regression (LR) method enhances the structural input by shell averaging, similar to the message passing of GNN. The SE(3)-GNN uses the equivariant message passing to learn more efficient geometric representation. The Glass-MLP extracts the Voronoi cell and energy from the inherent structure and uses the MLP with few parameters to predict dynamics. 
\begin{table}[]
\caption{Comparison to models that are designed to enhance geometric expressiveness. The results of linear regression (Linear), SE(3)-equivariant GNN (SE(3)-GNN), and multi-layer perception of glass (GLass-MLP) are taken from \cite{PaperEmanuele}, \cite{pezzicoli2022se3equivariant} and \cite{jung2022predicting}, respectively.  The \textbf{best} results at each timescale are marked in {\color{purple}{purple}}. The \textbf{second} ones are marked in {\color{green}{green}}.}\label{ad}

\begin{tabular}{c|ccccccccc}
\hline
\multirow{2}{*}{\bf{Model}} & \multicolumn{9}{c}{$S(\mathbf{q}, t)$} \\ \cline{2-10}                                                       
                       & $S=0.8$ & $S=0.7$ & $S=0.6$ & $S=0.5$ & $S=0.4$ & $S=1/e$ & $S=0.3$ & $S=0.2$ & $S=0.1$ \\ 
                       \hline
LR                 & -       & -       & 0.566   & 0.601   & 0.615   & 0.623   & 0.627   & 0.630   & 0.625   \\
GNN                    & 0.515   & 0.524   & 0.584   & 0.613   & 0.633   & 0.639   & 0.647   & 0.653   & 0.656   \\
SE(3)-GNN              & \color{green}\bf{0.736}   & \color{green}\bf{0.663}   &\color{green}\bf{ 0.638}   & \color{green}\bf{0.706}   & 0.713   & 0.725   & 0.725   & 0.709   & 0.713   \\
Glass-MLP              & 0.164   & 0.226   & 0.544   & 0.688   & \color{green}\bf{0.733}   & \color{green}\bf{0.757}   & \color{purple}\bf{0.775}   & \color{purple}\bf{0.777}   & \color{green}\bf{0.751}   \\
Geo-GNN                & \color{purple}\bf{0.798}   & \color{purple}\bf{0.708}   & \color{purple}\bf{0.747}   & \color{purple}\bf{0.760}   & \color{purple}\bf{0.778}   & \color{purple}\bf{0.774}   & \color{green}\bf{0.773}   & \color{green}\bf{0.770}  &\color{purple}\bf{ 0.755}  \\ \hline
\end{tabular}
\end{table}
The compassion results of all methods on the lowest temperature $T = 0.44$ are shown in Table \ref{ad}. Compared to these advanced models, our model is still the best in overall performance. As we can see, SE(3)-GNN \cite{pezzicoli2022se3equivariant} also outperforms the GNN by improving structural expression ability. In contrast to our rotation-invariant Geo-GNN, SE(3)-GNN learns the equivariant geometric embedding that can implicitly encode the angle or even more geometric information. Although SE(3)-GNN has a comprehensive capacity to extract features, it does not account for the smoothness of dynamics. Hence, its performance is inferior to that of Geo-GNN. At the prediction of long-time dynamics, Glass-MLP \cite{jung2022predicting} demonstrates the superior predictive ability to SE(3)-GNN. At $S(\mathbf{q},t)=0.3$ and $0.2$, it surpasses the Geo-GNN, but the difference is minor. This demonstrates that the authors utilized effective structural parameters, such as Voronoi cell and energy, for learning the smooth dynamics. Additionally, the Glass-MLP includes an auxiliary loss to enforce the spatial correlation of dynamics. This loss restricts the dynamics of nearby particles to be similar, which is the same concept as learning the smooth pattern of long-term dynamics. However, these structural parameters and the loss do not appear to work at the short-time dynamics with the non-smooth pattern. We hypothesize that a second cause of the improved performance is the authors' choice of the inherent structure over the initial configuration. Because the intrinsic structure eliminates thermal perturbations in the initial structure, the noise of the structural features can also be minimized, thereby facilitating performance enhancement. Recently, the authors in \cite{alkemade2023improving} also proposed to use the inherent structure and the cage structure to improve the correlation between the dynamic and structure. Here, We do not compare our method to that of their work. In actuality, it is unfair to directly compare our model with their work because the latter employs as input the distance from the initial particle configuration to the center of the cage, which already contains information about particle dynamics and thus yields superior prediction results. Nonetheless, it is possible to design a model to learn this dynamic information from the initial structure, thereby enhancing the prediction performance. We will investigate this in our future work. 

\begin{figure}[htb]
  \centering
  \includegraphics{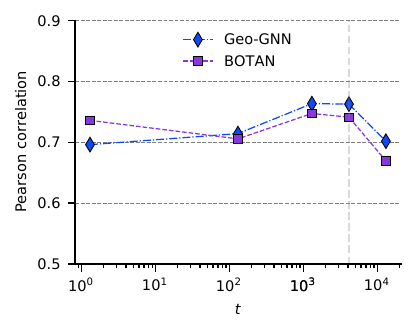}
  \caption{ \textbf{Comparison with BOTAN of predicting particle propensity in the KA system}. Pearson correlation coefficient between predicted and actual propensities for $A$ particles in the KA system at temperature $T = 0.44$. The vertical gray dashed lines indicate the timescale $t = \tau_g$, corresponding to $t=4120$. The data set and results are taken from Ref. \cite{shiba2023botan}. }
  \label{fig:botan}
\end{figure}

\textbf{Comparison to the BOTAN with improved aggregation strategy}

Recently, the authors in \cite{shiba2023botan} proposed learning the relative motion of particle pairs to alleviate the effect of elastic fluctuations on the prediction of dynamic propensities. They use one decoder layer on the last message aggregation to decode both the particle propensity on the node and the relative motion of particle pairs on edge. In fact, the relative motion of particles can be viewed as a physical constraint that prevents the propensity for particles to form smooth patterns. Thus, our method is essentially identical to learning relative motion directly, as both impose certain constraints on the space of features, making it easier to distinguish between nearby particles with distinct dynamics. The difference, however, is that their work uses an explicit constraint, while we implement an implicit one using the attention mechanism.

Here, we train our model on five selected time scales of the data set released in \cite{shiba2023botan} and report the results on in Fig.\ref{fig:botan}. Surprisingly, our method still exhibits impressive prediction performance without any additional dynamic constraints. Starting from the time point $t = 130$, our method gradually approximates BOTAN and surpasses it in the predictions around the relaxation time ($t= 4120$). The constraint on relative motion in aggregation does affect the prediction of short-time dynamics ($t=1.3$). However, Geo-GNN also achieves respectable results through aggregation under the attention mechanism with no additional dynamics. This result further demonstrates the efficacy of our method.

\begin{figure}[htb]
  \centering
  \includegraphics{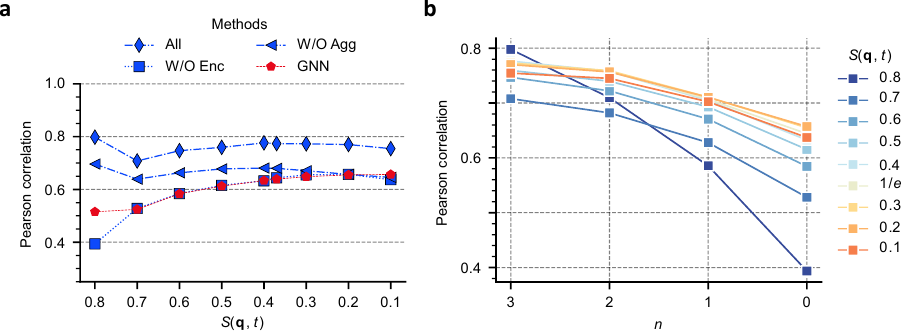}
  \caption{ \textbf{Ablation studies on the abilities of each component to predict particle propensity. }The results of the ablation study investigate \textbf{(a)} the effectiveness of the geometric feature encoder and feature aggregation and \textbf{(b)} the contributions of the triplet-to-edge layer in Geo-GNN.  We use the Pearson correlation to evaluate the predicted propensities of type $A$ at temperature $T = 0.44$. “W/O" is the abbreviation for 'Without'.}
  \label{fig:3a}
\end{figure}

\textbf{Propensity prediction ablation studies}

Our approach has successfully identified structure-dynamic relationships, providing a state-of-the-art method to enhance dynamics prediction in glassy systems. The GNN designs its feature vectors using several recurrent layers whose weights are shared across all stages. In contrast, we separate the geometric feature encoder from the feature aggregation and use a forward-learning technique to learn the features hierarchically. Moreover, we use high-frequency structural features and adaptive aggregation techniques to handle the smoothness of dynamics. Next, we examine the role of our proposed schemes in predicting propensity. As shown in Fig.~\ref{fig:3a}(a), the performance severely degrades when we drop the geometric feature encoder and keep only the aggregation process in the Geo-GNN (W/O Enc). However, such a simplified model behaves similarly to the GNN ones, with a slight difference only in the short timescale where $S(\bm{q}, t) = 0.8$. This indicates high-frequency structural representation plays a more important role in performance improvement, especially at short timescales. The GNN fails to capture such high-frequency features, leading to poor performance in predicting non-smooth dynamics. But the relative importance of the encoder is decreased in the longer timescales as the smooth dynamics rely less on the non-smooth structure feature. Furthermore, we replace the entire aggregation process with the MLP (W/O Agg), as done in the node update of the GNN model \cite{BAPST}, and the performance decrease at all timescales by directly aggregating the geometric features. These results demonstrate the efficacy of our proposed feature extraction and aggregation scheme. Incorporating information about the smoothness of dynamics is important for improving predictions of particle dynamics. 

Moreover, Fig.~\ref{fig:3a}(b) describes the performance of Geo-GNN as the function of the number of the triplet-to-edge layers in the encoder. It is possible to improve performance by using only one triplet-to-edge layer. The correlations can be improved by adding a second triplet-to-edge layer. But the inclusion of a third layer only slightly improves performance. Moreover, the model improves performance by increasing the number of layers at short timescales, especially in cage dynamics. As the receptive field of the encoder gradually expands, the differences between structural features will become evident due to the long-range disorder of structure. This discrepancy is more favorable to the formation of high-frequency features, and non-smooth dynamics can benefit from it. Thus, the dynamic propensity at short timescales with non-smooth patterns depends more on the depth of the encoder. Moreover, we prove that increasing the depth of the encoder can improve the ability to handle the smoothness of dynamics in the following experiments.

\definecolor{LightCyan}{rgb}{0.88,1,1}
\begin{table}[] 
\caption{Ablation study of the input features used to predict A-type particle propensities at temperature $T = 0.44$. With different input features, the {\bf{best}} results at each timescale are marked in {\color{purple}{purple}}, while the {\bf{worst}} ones are marked in {\color{blue}{blue}}.}\label{se}
\begin{tabular}{c|c|ccc}
\hline \multirow{2}{*}{\bf{Input feature}}  & \multirow{2}{*}{\bf{{Model}}} & \multicolumn{3}{c}{$S(\bm{q},t)$}    \\ \cline{3-5}

 &   & $S=0.8$ & $S=1/e$ & $S=0.1$ \\ \hline
\multirow{4}{*}{Distance and angle }  & Geo-GNN$^{(3)}$ & \color{purple}\bf{0.798} & \color{purple}\bf{0.774} & \color{purple}{\bf{0.755}} \\ 
& Geo-GNN$^{(2)}$ & 0.710   & 0.758   & 0.745   \\ 
& Geo-GNN$^{(1)}$ & 0.586   &  0.707  &  0.702  \\ 
& GNN        & \color{blue}0.515   & \color{blue}0.638   & \color{blue}0.658   \\ 
\hline
\multirow{4}{*}{Only distance } & Geo-GNN$^{(3)}$ & 0.408 &\color{purple}\bf{0.650} & \color{purple}\bf{0.679}  \\
 & Geo-GNN$^{(2)}$ & 0.402   & 0.644  & 0.676   \\ 
& Geo-GNN$^{(1)}$ & \color{blue}0.394   &  \color{blue}0.634  &  \color{blue}0.655 \\ 
& GNN  & \color{purple}\bf{0.413}   & 0.638   & 0.655   \\ \hline
 
\end{tabular}
\end{table}

One may now wonder whether the angular order or the distance affects the prediction performance of the model. Here, we only retain the distance information and the three-body connections at the input graph. Intriguingly, all revised models yield the same performance without angle features corresponding to triplet particles, as shown in Table \ref{se}. The increase in triplet-to-edge layers does not help improve accuracy in both short- and long-term predictions. Thus, the three-body connections accompanied by distance information can not provide sufficient structural details. It is also observed that knowing angle information for GNN only slightly improves the performance on short-time scales and achieves the same performance in predicting long-time dynamics by using distance as input. However, the Geo-GNN  can gain more performance improvement from the increase in the number of layers with the angle features, particularly in the case of the shorter-time dynamics. The impact of increasing the encoder layer still plays an important role in the longer-time predictions, although it is weakened. Thus, we believe that adding angle information is vital for improving the geometric expressive power of Geo-GNN.  

\begin{figure}[!htbp]
  \centering
  \includegraphics{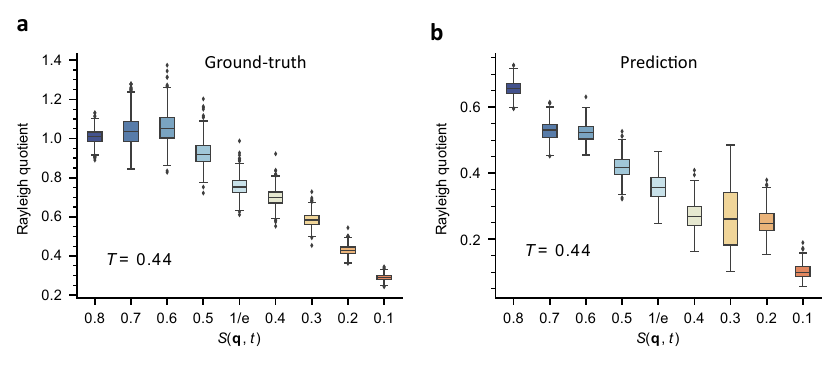}
  \caption{\textbf{Results on predicting global smoothness}. Rayleigh quotient of the \textbf{(a)} ground-truth propensities and \textbf{(b)} the predicted propensities for all test graphs at temperature $T = 0.44$. }
  \label{fig:re2}
\end{figure}

\begin{figure}[!htbp]
  \centering
  \includegraphics{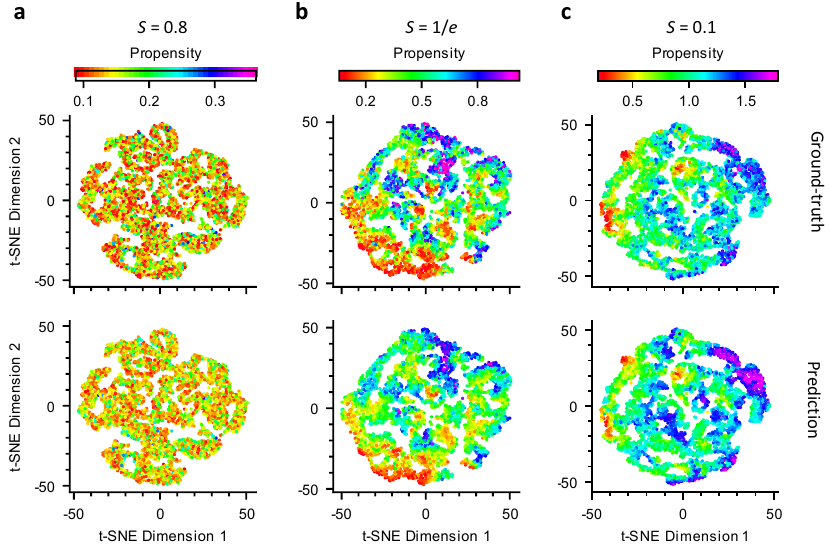}
  \caption{\textbf{Visual comparison of the predicted results}. \textbf{(a)-(c)} A two-dimensional visualization of the actual propensities (upper row) and Geo-GNN predicted (lower row) ones for particles of type $A$ at temperature $T = 0.44$. The three-dimensional positions under periodic boundary conditions are mapped to two-dimensional space by the t-SNE algorithm \cite{van2014accelerating} for visualization. }
  \label{fig:7vls}
\end{figure}

\textbf{Predicting global smoothness}

Till now,  we have verified the validity of the model in predicting the dynamics. Next, we deeply understand the ability of the model to handle the smoothness of dynamics, thus making the model more interpretable. Firstly, we examine the global smoothness expressed by Geo-GNN. Fig.~\ref{fig:re2}(a) and (b) describe the Rayleigh quotient of the ground truth and the predicted dynamics at $T = 0.44$ for all test graphs, respectively. The global smoothness of predicted dynamics shifts similar to the ground truth as time increases. This thoroughly explains the predicted dynamics that exhibit the high-frequency patterns on the short timescales and the low-frequency patterns on the long timescales. We show the target and predicted dynamic propensity in Fig.~\ref{fig:7vls}. For better visualization, we first transform the 3-dimensional positions to the 6-dimensional space to deal with the periodic boundary condition. Then, the t-SNE algorithm \cite{van2014accelerating} is used to project a 6-dimensional position of each particle into two-dimensional space. This scheme can better reveal the spatial relations of particle dynamics. Details of the visualization process can be found in Supplementary Note 4. We can see that the predicted propensities of our model agree well with the targets. The predicted dynamics between neighbors tend to be dissimilar at shorter timescales and become smooth at longer timescales, which is consistent with our actual dynamic behavior. Gaining access to these non-smooth patterns of dynamics allows our model to improve its accuracy on shorter-time predictions. The smoothness properties of diffusive dynamics require predictions that depend more on the shared characteristics of neighboring particles.

\begin{figure}[htb]
  \centering
  \includegraphics{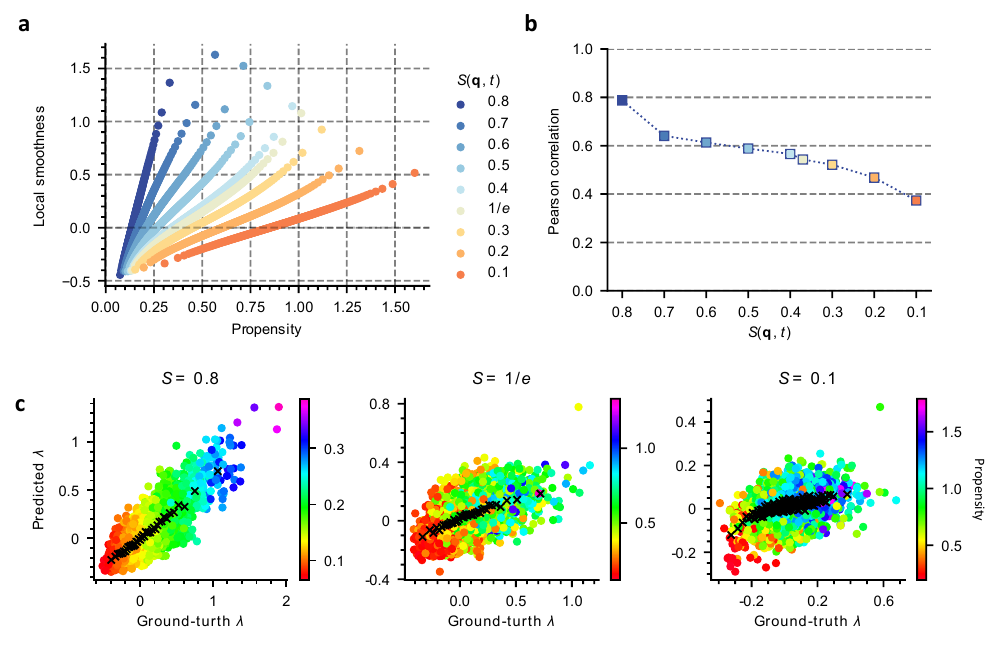}
  \caption{ \textbf{Experiment results on predicting local smoothness.} \textbf{(a)} Scatter plot between the actual and predicted propensity and local smoothness for particles of type $A$ at temperature $T = 0.44$. Each point at the plot is obtained by coarse-graining with a bin size of about 5000 particles \textbf{(b)} Person correlation between the actual and predicted propensity and local smoothness. \textbf{(c)} Scatter plot between the actual and predicted local smoothness in one configuration. Black markers are the results after coarse-graining with about 90 particles.}
  \label{fig:8}
\end{figure}

\textbf{Predicting local smoothness}

To further clarify the relationship between the dynamics and smoothness, we show in Fig.~\ref{fig:8}(a) the coarse-grained scatter plot between the propensities and local smoothness of all particles in the test data set. The local smoothness displays a strong correlation with the dynamic propensities. In the short-time scale where $S(\bm{q},t) = 0.8$, the relationship between local smoothness and propensities is nearly linear dependent. At the longer relaxation time, the correlation remains significant. This indicates that the dynamics in the environment always accompany the local smoothness of the particle and therefore is an inherent property of the particle dynamics. Based on this observation, we can further use the local smoothness to assess whether the dynamical environment governed by the central particle and its neighbors is well predicted. Fig.~\ref{fig:8}(b) displays the Pearson correlation between ground truth and predicted local smoothness. The high correlations further indicate that our approach can adaptively learn the underlying dynamic heterogeneity at various times. Here, we can see that our model captures the local smoothness of the particles very well at the short timescales. This can be further confirmed between the scatter plot of the ground truth and predicted local smoothness (Fig.~\ref{fig:8}). The correlation decreases as time increases. We suspect that the reason is that the local smoothness distribution gradually concentrates with increasing time, and the dynamics in the local domain gradually tend to smooth (Fig.~\ref{fig:g-l}(b)), resulting in a decrease of the discrimination in the local smoothness over long time scales compared to short time scales. This causes the prediction of dynamic smoothness to decrease over long time scales. However, as shown in Fig.~\ref{fig:8}(c), the predictions still distinguish well between particles with high and low smoothness in the longer timescales.

\begin{figure}[htb]
 \centering
  \includegraphics{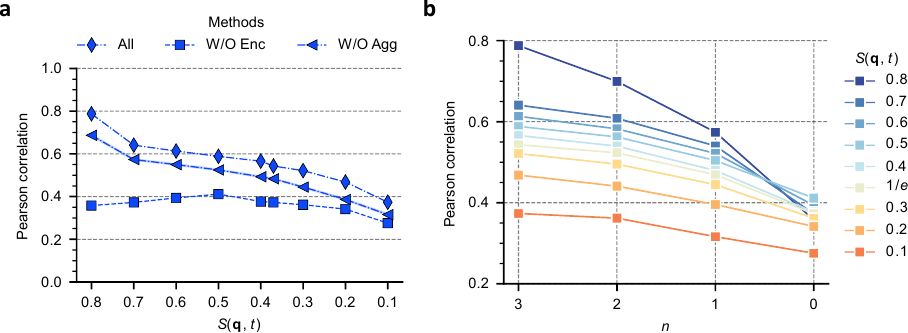}
  \caption{ \textbf{Ablation studies on the abilities of the encoder and aggregator to handle local smoothness. }The results of the ablation study were conducted to determine \textbf{(a)} the effectiveness of the geometric feature encoder and feature aggregation, as well as \textbf{(b)} the contributions of the triplet-to-edge layer in predicting local smoothness. The Pearson coefficient between the actual and predicted propensity particles of type $A$ at temperature $T = 0.44$ is used for the comparison.}
  \label{fig:9}
\end{figure}

\textbf{Local smoothness prediction ablation studies}

Next, we investigate the validity of each component in handling smoothness. As seen in Fig.~\ref{fig:9}(a), the correlation between the predicted and actual local smoothness decreases without either the encoder or the aggregator. We can see that both the encoder and aggregator affect the predictions of the local smoothness. The prediction results are consistent with the predictions of dynamic propensities (Fig.~\ref{fig:3a}(a)). Our proposed two-step aggregation can help us learn the local smoothness of dynamics more effectively than MLP. This means that averaging the messages by MLP on the edges directly can smooth out high-frequency information. Thus, it fails to predict the underlying smoothness of dynamics, leading to performance degradation. But our aggregator can explicitly put the smoothness constraints as the attention between neighbor particles. In this way, the features of the neighbor particle with the same targets will pull together to formulate low-frequency patterns in the local environment. On the other hand, features with distant targets will push away to form a high-frequency representation.

Moreover, we can also find that the model without the feature encoder layer performs poorly in capturing the smoothness of dynamics. Especially in the short time, the local smoothness of predicted dynamics deviates significantly from the ground truth, resulting in less accuracy in dynamic propensity prediction. Without the geometric encoder, high-frequency features can not be learned because the geometric expressive power is lost. The aggregator can not perform well for non-smooth dynamics with only the low-frequency features as input. Thus, the geometric encoder is important to provide the required structural factors for predicting dynamic smoothness. Partially at the short-time scales, we can gain more performance improvement on predicting short-time dynamics with the help of the high-frequency features that provide by a precise description of the local environment. These results demonstrate that our thought on characterizing the smoothness of particle dynamics by changing the structure extraction and aggregation blocks of the model does impact the performance for predicting dynamics.

\begin{figure}[htb]
 \centering
  \includegraphics{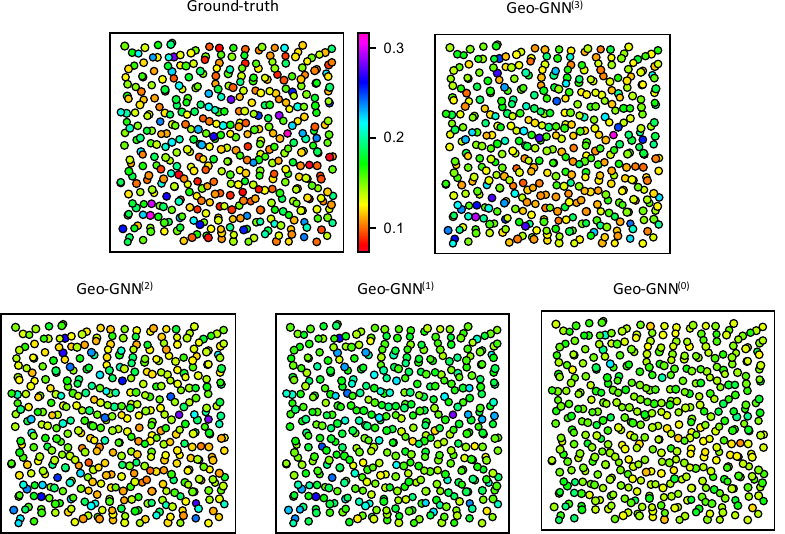}
  \caption{\textbf{Visualization of predicted propensity.} The plot of the predicted propensity from the Geo-GNN$^{(3)}$, Geo-GNN$^{(2)}$, Geo-GNN$^{(1)}$, and Geo-GNN$^{(0)}$ evaluated on one configuration at $S(\bm{q}, t) = 0.8$. Geo-GNN$^{(n)}$ refers to Geo-GNN using $n$ layers of triplet-to-edge message passing. The width of the shown slice is 2 $(11.04< z < 13.04)$. }
  \label{fig:vis}
\end{figure}

To further clarify the ability of the encoder to learn the smoothness of dynamics, we investigated whether the triplet-to-edge layer depth affected the model's capacity to capture the smoothness of dynamics. In Fig.~\ref{fig:9}(b), we report the Pearson correlation between the actual and predicted local smoothness as a function of the number of triplet-to-edge layers. Consistent with the performance of dynamics predictions, the correlations for local smoothness increase as the model accesses longer-range orientation order. Notably, even a single layer significantly enhances the ability to learn the smoothness of dynamic heterogeneity and improve the accuracy of the predictions. Particularly at the short-time scale with non-smooth dynamics, the depth of the geometric feature encoder is crucial for capturing spatial heterogeneity at these times. The results can be confirmed from the plots in Fig.~\ref{fig:vis}. As the encoder layer decreases, the dynamic variations of neighboring particles are gradually lost. Finally, the predicted propensity of the Geo-GNN$^{(0)}$ makes the dynamics smooth in the system. Thus, the long-range dependencies allow the model to learn more distinguishable features vital for dynamic smoothness. These results again demonstrated that improving representation learning of high-frequency features can better reveal the underlying smoothness. Also, this interprets why the geometry feature extraction phase is important for predicting glassy dynamics.

\section*{Conclusion}

In this paper, we use graph signal processing tools to answer the question of which type of structure is required to predict the dynamics of a glass system. In contrast to previous studies, we show that smoothness is an intrinsic feature of glassy dynamics that differs sharply between long and short timescales. The relevant structural features should have consistent smoothness constraints with the dynamics when building a connection between structure and dynamics.  That is, accurate prediction of short-time dynamics with non-smoothness nature depends more on high-frequency structural features. In contrast, accurate prediction of long-time dynamics relies more on low-frequency structural features. This provides new clues for future research on the structural origin of dynamical heterogeneity. Based on dynamical smoothness, we illustrate that two approaches in learning the structure can enhance the performance of dynamic predictions: 1) by boosting the structural representation ability of the local environment to avoid indistinguishable between near-neighbor particles, and 2) by improving aggregation operations that can constraint the structural representation with the smoothness of dynamics. Combining these approaches can provide a novel framework for improving prediction performance.

We next proposed a geometrically enhanced graph neural network model based on these observations with improved accuracy and robustness in predicting dynamics. Specifically, the model uses a geometry-informed feature encoder module with high-order message passing to enrich the structural expressive power of the local environment. An adaptive aggregation strategy that learns local smoothness implicitly is then proposed to constrain the learned geometric structures based on the smoothness pattern of the neighbor particle. Finally, the proposed model outperforms the current methods in predicting the dynamic propensity of the 3D Kob-Anderson mixture. 

Our experiment results show that the proposed geometry-informed feature encoder of the local environment can significantly enhance the expressiveness of smoothness. Moreover, adaptive aggregation of the learned geometric features can also strengthen the ability to understand dynamic smoothness. Furthermore, increasing the number of geometric feature encoder layers improves the structural feature representation and the ability to capture smoothness in dynamics. The success of our proposed model demonstrates our hypothesis that prediction performance can be enhanced by improving its ability to learn the underlying smoothness. This viewpoint provides a foundation for future research on the structure-dynamics relationship. 

In summary, our study demonstrates that predicting dynamics at different times requires structural features with different smoothness. Although the suggested model provides robust prediction, it is still less effective at predicting short-time dynamics, particularly in the cage stage. However, we can still advance our method under the concept of learning dynamical smoothness. For instance, we can further enhance the geometric expressive power by achieving a completeness representation of particle structure \cite{ bartok2013representing, pozdnyakov2020incompleteness, nigam2023completeness} or by introducing inherent structure \cite{jung2022predicting} and cage structure \cite{alkemade2023improving} into the initial configurations. Consequently, the ability to differentiate between arbitrary configurations allows for generating more high-frequency representations corresponding to the non-smooth cage dynamics. In addition, we can further constrain the model in feature aggregation operations by substituting the implicit constraint on smoothness with an explicit one, for example, by incorporating the relative motion or more high-order dynamic constraint between particles as an inductive bias \cite{shiba2023botan}. Combining these methods can further improve the correlation between dynamics and structure. However, integrating this knowledge necessitates a new version of the model, which we leave in future studies.

\section*{Methods}

\textbf{Graph signal and smoothness} 

An undirected graph $\mathcal{G}=(\mathcal{V},\mathcal{E})$ can be used to represent a single static snapshot of the glassy system, where $\mathcal{V}$ is the set of nodes that correspond to particles, and $\mathcal{E}$ is the edge set that reveals the connections between the nodes. An undirected edge links two nodes if their distance is smaller than a defined threshold. Here, we only consider the particles of type $A$ when constructing the graph. The graph can be further described by an adjacency matrix $\bm{A} \in{\{0,1\}^{N_A\times{N_A}}}$, where $A_{ij} = 1$ indicates that a connection exists from $i$ to $j$ in the graph (i.e., $(i,j)\in{\mathcal{E}}$) and $A_{ij} = 0$ otherwise. Since $\bm{A}$ is unweighted, we have $A_{ij} = A_{ji}$. The representation of the graph Laplacian can be obtained by $\bm{L} = \bm{D}-\bm{A}$, where $\bm{D}$ is the degree matrix with the diagonal elements $d_{ii} =\sum_{j = 1}^{N_A} A_{ij} $ and $d_{ij} = 0$ for $i \neq j$. 

Each node $i$ in a built graph can be assigned with its dynamic propensity $\Delta_i(t)$ at time $t$. A vector $\bm{\Delta}(t) = [\Delta_1(t), ...,\Delta_{N_A}(t)]$ that stacks these scalar values is called a graph signal. The Rayleigh quotient of a graph signal $\bm{\Delta}(t)$ supported on the graph is defined as

\begin{equation}
        R(\bm{L}, \bm{\Delta}(t)) = \frac{\bm{\Delta}(t)^{T}\bm{L}\bm{\Delta}(t)}{\bm{\Delta}(t)^{T}\bm{\Delta}(t)} = \frac{\sum_{i\in{\mathcal{V}}}\sum_{j\in{\mathcal{N}_i}}(\Delta_i(t) - \Delta_j(t))^2}{\sum_{i\in{\mathcal{V}}}\Delta_i^2(t)},
\end{equation}

where $\mathcal{N}_i$ is the neighbor set of particle $i$. The Rayleigh quotient can be considered a way to measure how smooth a signal is on the graph. A higher Rayleigh quotient value indicates that the graph signal fluctuates strongly across the nodes, corresponding to the high-frequency graph signals. The Rayleigh quotient has also been used to understand graph neural networks from the graph filtering perspective \cite{hoang2021revisiting}. 

To quantify the local smoothness at the particle level, we also defined the local propensity deviation for each node as follows:
\begin{equation}
       \lambda_i(t) =\frac{1}{d_{ii}}\sum_{j\in{\mathcal{V}(i)}}\frac{\Delta_i(t) - \Delta_j(t)} {\Delta_j(t)}.
\end{equation}
The local propensity deviation measures how much the signal values at central node $i$ differ from neighboring nodes. The larger the absolute value of local propensity deviation, the more significant the difference between the central particle and its surrounding neighbors.

\textbf{Geo-GNN model}

\emph{Graph input layer} 
Following~\cite{BAPST}, we first establish a basis graph consisting of the node set $\mathcal{V}$ and edge set $\mathcal{E}$ from the input 3D particle glass system. Each particle is regarded as a node $i\in\mathcal{V}$. Two directional edges $(i, j)$ and $(j, i)$ connect a pair of nodes $i$ and $j$ with opposite directions when the interparticle distance $d_{ij} = \lVert\bm{r}_i - \bm{r}_j\rVert$ is less than the cutoff threshold. We choose the threshold as 2.0 in the main text. The result is almost unaffected by the threshold when greater than 1.25 (see Supplementary Note 1). Since the primary relative positions are changeable and inconsistent across various coordinate systems, we consider the rotational-invariant distance as the edge feature providing the two-body geometry information. Then, we have two directed edges with the same feature as $d_{ij} = d_{ji}$ but are in opposing directions. Moreover, we add the type $Z_{ij}$ for edge $(i,j)$ based on the particle types of nodes $i$ and $j$. For the binary system with particle types $A$ and $B$, there will be four types of edges labeled $0$, $1$, $2$, and $3$, which correspond to $A\to{A}$, $A\to{B}$, $B\to{A}$, and $B\to{B}$, respectively.

In contrast to the present GNN, which only includes two-body connections, we start with the basic graph and include three-body connections $\mathcal{A}$ corresponding to triplets of particles. Given pair of edges $(i,j)$ and $(j,k)$ on the basis graph, the three-body connections $(i, j, k) \in \mathcal{A}$ with $j\ne i$, are formed when the target node $i\in \mathcal{N}(j)$ interacts with the source $k\in \mathcal{N}(j)$ by the passing node $j$. Then, angles $\theta_{ijk}= argcos(\frac{\bm{r}_{ij}\bm{r}_{jk}}{\lVert\bm{r}_{ij}\rVert\lVert\bm{r}_{jk}\rVert})$ between two edges can be explicitly incorporated into the graph as the initial triplet features. Ref.~\cite{BAPST} considered the three-dimensional relative positions and assigned them as the input graph's edge features. However, when the relative position of two particles in Euclidean space is projected onto a high-dimensional space, the original distance information is well-kept, but the particle's angular geometry in Euclidean space may be hidden. Moreover, the sensitivity of the coordination makes GNN susceptible to rotation. As a result, the model necessitates additional online data augmentation to enhance generalization. Unfortunately, augmentation creates many scattered 3D positions, making it harder to learn about local structures from geometric features. Instead of using relative locations and distances, our model may leverage angle information to describe the local structure of particles in more detail.

\emph{Graph embedding layer}
Before the start of the geometric encoding, the scalar properties (i.e., edge type, distance, and angle) are first encoded to get initial scalar embeddings for all edges and triplets. Here, we do not use any initial features tied to individual nodes. The embedding layer first converts types of edge to their one-hot representations and further embeds them into dense vectors $\bm{z}_{ij}$:
\begin{equation}
    \bm{z}_{ij} =\bm{1}_{HoT}(Z_{ij})\bm{W_z},
\end{equation}
where $\bm{1}_{HoT}( )$ is the one-hot encoder of $Z_{i,j}$, and $\bm{W}_z$ is the trainable embedding matrix initialized randomly and optimized in the training process. Next, we adopt the radial basis functions layer to encode the interparticle distances: $\bm{e}_{ij} = [R_1(r_{ij})\lVert{}… \lVert{}R_{N_{r}}(r_{ij})]$, where $\lVert{}$ represents concatenation over the scalars to get ${N_r}$-dimensional embedding, $R_{n_r}(r_{ij}) $ is the $n_r$-th radial basis function. Here, we use the Gaussian radial basis function $R_k(r_{ij}) = e^{-\beta(r_{ij}-\mu_{k})}$, with $\mu_{k}$ being uniformly selected between zero and the distance cutoff $R_c$, while $\beta$ is a learnable scaling parameter. The encoded type and distance can be further combined to get the initial latent features for each edge:
\begin{equation}
    \bm{h}^0_{ij} = MLP([\bm{z}_{ij}\lVert{\bm{e}_{ij}}]).
\end{equation}

For the initial triplet features, we use the projection of spherical harmonics on $\theta_{ijk}$ to encode the three-body correlations, which are given by:
\begin{equation}
    \bm{t}_{ijk} = [Y^0_1(\mathbf{\theta}_{ijk})\lVert … \rVert Y^0_{N_{\theta}}(\mathbf{\theta}_{ijk})]\bm{W}_{\theta},
\end{equation}
where $Y^0_l(\mathbf{\theta}_{ijk})$ is the $l$-th order real spherical harmonic, and $\bm{W}_{\theta}$ is the trainable parameters. Empirically, these representations can provide detailed geometry information to the network and help our model converge fast during training. 

\emph{Geometric feature encoder}
After the initial embeddings are established, Geo-GNN conducts angle-aware convolutions in the triplet-to-edge layer to comprehend the geometric environment. The rules of the convolution that propagate the features from the triplet to the edge can be referred to as the message passing on a line graph \cite{Gasteiger2020Directional,choudhary2021atomistic,fang2022geometry}. As shown in Fig.~\ref{fig:model}(d), the embedded environment for edge $Z_{i,j}$ is updated by employing the following angle-informed convolution:

\begin{equation}
    \bm{m}_{ij}^{L} = \sum_{k\in{\mathcal{N}_{j}\setminus{\{i\}}}}\bm{W}^L_{\theta}\bm{t}_{ijk} \odot \bm{W}^L_{e}\bm{h}^{L-1}_{jk}
    \label{eq_edge}
\end{equation}
\begin{equation}
	\bm{h}_{ij}^{L} = \sigma(\bm{W}^L_{e'}[\sigma(\bm{W}^L_{e''}\bm{h}^{L-1}_{ij}) \lVert{} \bm{m}^{L}_{ij}]).
    \label{eq_edge1}
\end{equation}
where $\sigma$ is the activation function, $\odot$ denotes the element-wise multiplication and $\bm{W}^L_{\theta}$, $\bm{W}^L_{e}$,$\bm{W}^L_{e'}$  and $\bm{W}^L_{e''}$ are trainable weight matrix. Unless otherwise stated, Swish/Sigmoid Linear Unit (SiLU) activation functions are used throughout the text since they may improve empirical performance on various tasks. After performing $L$ triplet-to-edge layers, information from multi-hops away can be incorporated into the learned geometric descriptors associated with pairs of neighboring particles. To allow the model easily propagate the features from earlier layers forward, we use the residual update by concatenating the newly updated geometric information with the features from the previous $L-1$ layer to yield the final geometric environment embeddings as in Eq.~\ref{eq_edge1}. 

Traditional methods typically rely on hand-crafted descriptors such as symmetry functions \cite{behler2007generalized} to describe the particle environment in a rotationally invariant way. The radial symmetry function is used to characterize the radial density environment, while the angular symmetry function is used to characterize the bond orientations. These methods first project the distance and angle between the central and neighbor particles onto a combination of basis functions. The sum of these project features of neighbor particles is then used to describe the local environment of the central particle. Naturally, this method results in coarse-grained descriptors, which fail to capture the essential details of dynamic smoothness. Here, we characterize the local environment for all edges instead of nodes to achieve a more fine-grained description. As shown in Eq.~\ref{eq_edge}, angle-aware convolution can be viewed as a weighted sum of the bond orientations in the local environment. But unlike hand-crafted approaches, we can learn the weights of the sum from the neural network. Then the embedded environment descriptor $\bm{m}_{ij}^{L}$ can contain more expressive three-body information.

\emph{Feature aggregation} 
When the geometric environment features for two nearby particles are extracted, we aggregate the structural features at the node level by using two edge-to-node layers and one node-to-node layer. According to the property of dynamic smoothness, the structural features after aggregation should reflect the underlying smoothness that correlates to dynamic heterogeneity. Therefore, beyond averaging the local structure, we design a geometry-enhanced self-attention mechanism to assign different attention weights between nodes that can adaptively handle the propagation of high- and low-frequency structural features. 

In the first step of edge-to-node aggregation, we initialize the message with a skip connection from the initial edge embeddings $\bm{h}^0_{ij} $  to this layer for keeping more original two-body geometric information. Next, we augment the edge message from node $i$ to $j$ with the learned high-order geometric details by concatenating the newly updated node features $\bm{h}^0_i$ with environment embeddings $\bm{h}^L_{ij}$ learned by the geometric feature encoder. Specifically, two layers of edge-to-node message passing are given:
\begin{equation}
	\bm{\alpha}_{ij} = tanh(\bm{W}_{\alpha}\bm{e}_{ij}))
\end{equation}
\begin{equation}
\bm{h}_i^0 = \sum_{j\in\mathcal{N}_i}\bm{\alpha}_{ij}\odot \bm{W}^0\bm{h}^0_{ij}
\end{equation}
\begin{equation}
	\bm{\alpha}'_{ij} = tanh(\bm{W}_{\alpha'}[\bm{h}_{i}^0\lVert\bm{h}^L_{ij}\rVert\bm{h}_{j}^0])
\end{equation}
\begin{equation}
\bm{h}_i = \sigma(\sum_{j\in\mathcal{N}_i}\bm{\alpha}'_{ij}\odot \sigma(\bm{W}^1 [\bm{h}_{i}^0\lVert\bm{h}^L_{ij}\rVert\bm{h}_{j}^0])),
\end{equation}
where $\bm{W}_{\alpha}$, $\bm{W}_{\alpha}'$, $\bm{W}^{0}$, and $\bm{W}^{1}$ are trainable weight matrices, and $tanh$ is the hyperbolic tangent activation function whose output is ranged in $(-1,1)$. Here, we propose a geometric-enhanced self-attention strategy where the importance of the neighbor nodes is calculated from the learned geometric feature. Naturally, the neighbor nodes that received a highly important score contribute more to the aggregation of the target node feature. Traditional attention schemes, such as graph attention neural network (GAT) \cite{velivckovic2017graph}, typically use the softmax to force the attention coefficients to be positive. But the aggregation based on non-negative attention scores is still equivalent to averaging the local structure, which remains unfavorable for learning high-frequency structural features. By relaxing the attention weights to negative values through the $tanh$ activation function, the target node feature can be augmented not only by the similarity with the non-negative attentions but also by the differences of neighboring nodes with negative attention. Thus, we use the neighboring difference to learn more high-frequency dynamic patterns beyond averaging the local structure. Similar attention strategies have also been unitized in other domains \cite{bo2021beyond, yang2021diverse}.

To provide more accurate predictions, we further fine-tune the node feature to enhance the high-frequency components from its neighboring environment rather than conducting the node regression directly. The following node-to-node aggregation is performed:
\begin{equation}
\bm{h}_i^{out} = \sum_{j\in\mathcal{N}_i}MLP([\bm{h}_{i} \lVert \bm{h}_{i} - \bm{h}_{j}]).
\end{equation}
This layer explores the dissimilarity between a node's self-feature and its local environment. Intuitively, the aggregation of the difference between nodes works similarly to a high-pass filter, enabling it to learn more high-frequency information and preventing the node from being homogeneous. Combining the self-feature with the neighbor difference allows the model to understand the crucial component from the supervised dynamic propensity adaptively.

\textbf{Training details}

The loss function is calculated by minimizing the L2-norm of the difference between the predicted and actual probabilities of type $A$ particles. The architecture and hyper-parameters were optimized at the temperature $T = 0.44$ and $S(\bm{q},t) = 1/e$. The Geo-GNN is implemented by three layers of triplet-to-edge for the geometric feature encoder, two layers of edge-to-node with 32 hidden units, and one layer of node-to-node, each having 32 hidden units. The cutoff and number of the radial basis function as $5.0$ and $N_r = 64$, respectively. We use $N_\theta = 16$ for the representation of the angle. In all our tests, we train Geo-GNN using the Adam optimizer \cite{adam} with an initial learning rate of 0.001.

\textbf{Data availability} 

The MD simulation trajectories of the Kob-Anderson system are available in Ref \cite{BAPST}.

\textbf{Code availability} 

Geo-GNN is implemented using PyTorch Geometric \cite{Fey/Lenssen/2019}. The code is available from \url{https://github.com/xjiang-hnu/Geo-gnn}.

%

\section*{Additional information}

\textbf{Acknowledgements} We thank  V. Bapst for sharing the dataset and F. P. Landes for providing the results. This work was financially supported by the Natural Science Foundation of Hunan Province (Nos.2022JJ30166).

\textbf{Author contributions.} X.J. developed the software and performed the analysis. X.J., Z.A.T., and K.L.L. contributed to the interpretation of the results. X.J. and Z.A.T. conceived the idea and approach presented in this work. All authors contributed to the writing of the paper.

\textbf{Competing interests.} The authors declare no competing interests.

\textbf{Materials \& Correspondence.} Correspondence and requests for materials should be addressed to Z.A.T. (tianzean@hnu.edu.cn). 

\widetext
\clearpage

\setcounter{equation}{0}
\setcounter{figure}{0}
\setcounter{table}{0}
\setcounter{page}{1}
\makeatletter
\renewcommand{\figurename}{Supplementary Figure}
\renewcommand{\tablename}{Supplementary Table}

\begin{center}
\textbf{\large Supplementary Information: Geometry-enhanced graph neural network for learning the smoothness of glassy dynamics from static structure}
\end{center}

\begin{center}
Jiang et al.
\end{center}
\textbf{Supplementary Note 1: impact of edge threshold.} 

In this note, we first explore how the performance of our proposed geo-GNN depends on the edge threshold employed in the graph's construction. To this end, we train Geo-GNN at $S(\bm{q},t) = 0.8$, $1/e$, and $0.1$ using the graphs with different edge thresholds as the inputs. As shown in Supplementary Fig. \ref{fig:s1}, when the edge threshold is greater than $1.25$, the performance of our model is weakly affected by the edge threshold, especially for the long-time dynamics. As for the short timescale, the accuracy slightly improves when the threshold increases. Even though we achieve the greatest performance with an edge threshold of $2.0$ (see Supplementary Table.\ref{tab:edge}). However, the graph with an edge threshold of $1.5$ may reduce the number of triplets by over an order of magnitude, resulting in significant computation time savings.
\begin{figure}[htbp]
  \centering
  \includegraphics{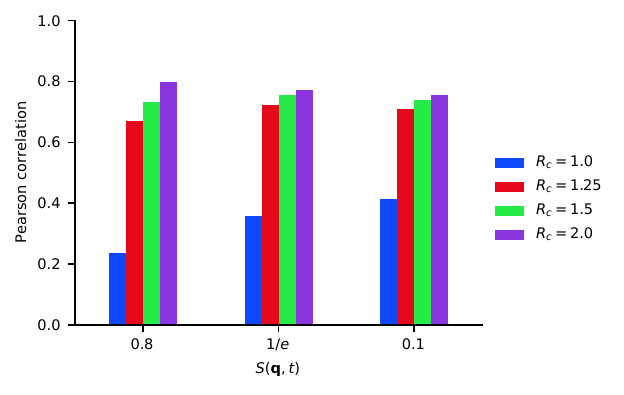}
    \caption{The Pearson correlation between the dynamic propensity predicted by Geo-GNN and ground-truth as a function of edge threshold at temperature $T=0.44$. }
  \label{fig:s1}
\end{figure}

\begin{table}[h]
\caption{The number of edges and triplets with different edge thresholds. \label{tab:edge}}
\begin{ruledtabular}
\begin{tabular}{ccccc}
Edge threshold     & 1.0                & 1.25               & 1.5                 & 2.0                  \\
number of edges    & $\backsimeq$ 11800 & $\backsimeq$ 47000 & $\backsimeq$ 61600  & $\backsimeq$ 167000  \\
number of triplets & $\backsimeq$ 43000 & $\backsimeq$ 510000 & $\backsimeq$ 875000 & $\backsimeq$ 6677000
\end{tabular}
\end{ruledtabular}
\end{table}
\clearpage

\textbf{Supplementary Note 2: parameter details.} 

We optimized the parameters of Geo-GNN at the state point where $T = 0.44$ and $t$ is chosen at $S(\bm{q},t) = 1/e$.  In order to balance performance and efficiency, we train the Geo-GNN on graphs with an edge threshold of $\bm{1.5}$ for selecting the hyper-parameters. The parameter search space and final parameters for the Geo-GNN architecture are specified in Supplementary Table.~\ref{parameter}. The final parameters are marked in \textbf{bold}. 
The detailed comparison results of different dimensions of the hidden layers and the length of the radial basis function are shown in Supplementary Fig.~\ref{hidden} and Supplementary Fig.~\ref{radial}. We can find that the results have little dependence on these parameters. It is possible to train the networks to get comparable results with a hidden dimension of just 16. Moreover, we also replace the Gaussian expanded representations of distance with a linear encoder (labeled as “1” at the horizontal coordinate of Supplementary Fig.~\ref{radial} ). It can be seen that there is no difference in the results.

Instead of using spherical harmonics to represent the angle, we also utilize the linear layer to encode the angle into a vector representation for dynamic prediction. As shown in Supplementary Fig.~\ref{angle}, the spherical harmonics representation of the angle improves the performance for all timescales, especially for the short timescales. Therefore, such a representation of angle may offer more reliable information on structural order than just utilizing angle alone.
\begin{table}[h]
\caption{The parameter search space for Geo-GNN at the temperature $T = 0.44$ and $S(\bm{q},t) = 1/e$}.\label{parameter}
\begin{ruledtabular}
\begin{tabular}{cc}
\hline
hidden dimensions              & {[}16, \textbf{32}, 64{]} \\
length of radial basis function & {[}16, \textbf{64}, 128{]} \\
length of spherical harmonics  & 16              \\
number of triplet-to-edge layer          & {[}1, 2, \textbf{3}{]}     \\
number of edge-to-node layer            & 2               \\
number of node-to-node layer           & 1               \\
dropout                        & {[}\textbf{0}, 0.1, 0.2{]} \\
batch size                       & 1              
\end{tabular}
\end{ruledtabular}
\end{table}

\begin{figure}[htbp]
  \centering
  \includegraphics{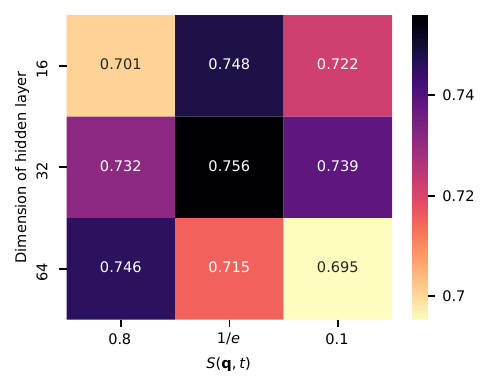}
    \caption{Ablation study to determine the dimensions of the hidden layer.}
  \label{hidden}
\end{figure}

\begin{figure}[htbp]
  \centering
  \includegraphics{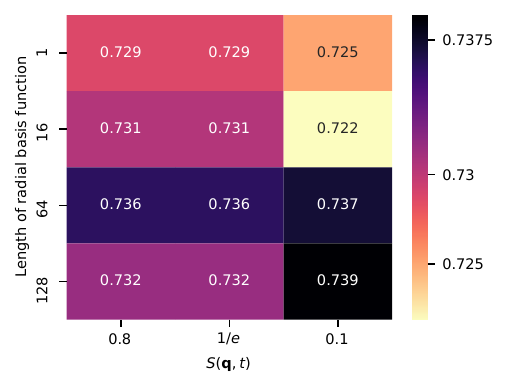}
    \caption{Ablation study to determine the length of radial basis function. }

  \label{radial}
\end{figure}

\begin{figure}[htbp]
  \centering
  \includegraphics{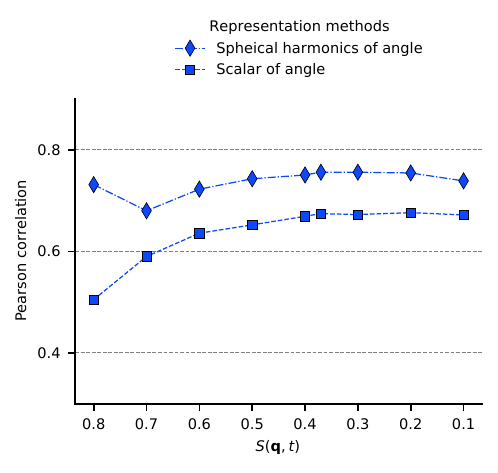}
    \caption{Ablation study to determine the representation strategy of angle. }

  \label{angle}
\end{figure}

\clearpage
\textbf{Supplementary Note 3: the detailed comparison between models.} 

Supplementary Table.~\ref{details} illustrates the details of Geo-GNN and GNN.

\begin{table}[h]
\caption{The detailed comparison between the Geo-GNN and GNN.}\label{details}
\begin{ruledtabular}
\begin{tabular}{ccc}
Model                     & Geo-GNN             & GNN                   \\
node size                 & 3277                & 3277                  \\
node feature size         & -                   & 1 (type)             \\
edge size                 & $\backsimeq$ 167000  & $\backsimeq$ 167000    \\
input edge feature size         & 2 (type, distance) & 3 (relative position) \\
triplet size              & $\backsimeq$ 6677000 & -                     \\
input triplet feature size      & 1 (angle)          & -                     \\
data augmentation         & without             & with                  \\
rotation-invariant        & True                & False                 \\
recurrent message passing & False               & True                  \\
number of parameters      & 30274               & 70721                
\end{tabular}
\end{ruledtabular}
\end{table}
\clearpage
\textbf{Supplementary Note 4: details of visualization.} 

To better show the spatial relations of the particle propensity, we project the 3-dimensional positions of each particle to 2-dimensional space under the periodic boundary conditions. To impose the periodicity along the x, y, or z directions, we map the position component $p$ in each direction onto a circle with a unit radius by using the following function:
\begin{equation}
    f(p) = [sin(\frac{2{\pi}p}{L}), cos(\frac{2{\pi}p}{L})],
\end{equation}
where $L$ is the cell length. Then, the 3-dimensional position of a particle can be transformed into the following 6-dimensional vector:
\begin{equation}
    [p_x, p_y, p_z] \xrightarrow{f(p)}{}  [sin(\frac{2{\pi}p_x}{L}), cos(\frac{2{\pi}p_x}{L}), sin(\frac{2{\pi}p_y}{L}), cos(\frac{2{\pi}p_y}{L}),sin(\frac{2{\pi}p_z}{L}), cos(\frac{2{\pi}p_z}{L})].
\end{equation}
These transformed positions keep the nearest neighbor relationship between particles in the original space. Next, the t-SNE algorithm can be applied to perform dimensionality reduction to keep the distance information. Finally, the results can be visualized in the 2-dimensional space. In this way, we can better examine the spatial correlation of predicted dynamics than the usually used contour plot. 

\clearpage

\end{document}